\documentclass[5p]{elsarticle}

\usepackage{multirow}

\usepackage{soul}
\usepackage{color}
	\definecolor{celadon}{rgb}{0.67, 0.88, 0.69}
    \definecolor{flamingopink}{rgb}{0.99, 0.56, 0.67}

\usepackage[table,xcdraw]{xcolor}
\soulregister\cite7

\usepackage{graphicx}

\usepackage{lineno,hyperref}
\modulolinenumbers[5]

\journal{Elsevier Digital Investigation}

\begin{document}

\begin{frontmatter}


\title{A Survey of Electromagnetic Side-Channel Attacks and \\Discussion on their Case-Progressing Potential for Digital Forensics}

\author{Asanka Sayakkara~\corref{mycorrespondingauthor}}
\ead{asanka.sayakkara@ucdconnect.ie}
\author{Nhien-An Le-Khac}
\ead{an.lekhac@ucd.ie}
\author{Mark Scanlon}
\ead{mark.scanlon@ucd.ie}
\address{Forensics and Security Research Group, University College Dublin, Ireland}

\begin{abstract}

The increasing prevalence of Internet of Things (IoT) devices has made it inevitable that their pertinence to digital forensic investigations will increase into the foreseeable future. These devices produced by various vendors often posses limited standard interfaces for communication, such as USB ports or WiFi/Bluetooth wireless interfaces. Meanwhile, with an increasing mainstream focus on the security and privacy of user data, built-in encryption is becoming commonplace in consumer-level computing devices, and IoT devices are no exception. Under these circumstances, a significant challenge is presented to digital forensic investigations where data from IoT devices needs to be analysed. 

This work explores the electromagnetic (EM) side-channel analysis literature for the purpose of assisting digital forensic investigations on IoT devices. EM side-channel analysis is a technique where unintentional electromagnetic emissions are used for eavesdropping on the operations and data handling of computing devices. The non-intrusive nature of EM side-channel approaches makes it a viable option to assist digital forensic investigations as these attacks require, and must result in, no modification to the target device. The literature on various EM side-channel analysis attack techniques are discussed -- selected on the basis of their applicability in IoT device investigation scenarios. The insight gained from the background study is used to identify promising future applications of the technique for digital forensic analysis on IoT devices -- potentially progressing a wide variety of currently hindered digital investigations.

\end{abstract}


\end{frontmatter}


\section{Introduction}
\label{intro}

Digital forensics is the field where legal investigations are assisted by analysing digital sources of evidence. In contrast, cybersecurity is the domain where the concern is to ensure the security of digital data and the privacy of their owners. In today's modern world, technology is becoming increasingly prevalent in everyday life and many people stay almost always connected to the Internet~\cite{nie2000internet}. While various social networks facilitate their users to share their life events to the rest of the world intentionally, every computer-based device they interact with in everyday life leaves unintentional traces of their activities. Such sources of forensic information include computer hard disks, network activity logs, removable media, internal storage of mobile phones and many others~\cite{soltani2017survey}.

Internet of Things (IoT) is an emerging trend started as a narrow research domain called wireless sensor networks, which evolved into Internet-connected everyday objects. IoT ecosystem includes a wide variety of devices, such as smart-watches, smart TVs, CCTV cameras, medical implants, fitness wearables, etc. The increasing availability of IoT devices across society makes it inevitable to find them in modern crime scenes and digital forensic investigations. Most of these devices comes with limited data processing and storage capabilities and they usually possess limited standard interfaces to the outside world, such as USB ports or WiFi/Bluetooth wireless interfaces, unlike their PC counterparts~\cite{stojkoska2017review}.

Due to the increasing concerns regarding security and privacy among communities, modern digital devices, such as computer systems, mobile devices, etc., are designed and shipped with built-in security.
Popular smartphones, such as iOS and Android based devices, encrypt their internal storage in order to protect user data from third parties~\cite{ahmad2013comparison}. Each of the mainstream PC operating systems, such as Mac OS, Windows, and Linux, provide built-in hard disk encryption. Meanwhile, network communications, both wired and wireless, commonly employ strong packet encryption mechanisms~\cite{vanDeWiel2018network}. Modern computer hardware has made the automated handling of encrypted data an everyday possibility in consumer, industrial and military applications~\cite{fritzke2012obfuscating}. Computer devices seized at a crime scene containing encrypted data poses a significant challenge to the investigation~\cite{lillis2016challenges,sayakkara2018EMForensics}. The IoT device ecosystem is no exception for this data encryption trend making the challenge of digital forensic investigations on IoT devices even more complex.

Side-channel analysis attacks have been proven to be useful to breach security on computer systems when standard interfaces, e.g., network interfaces and data storage devices, are sufficiently protected~\cite{spreitzer2018systematic, dhem1998practical, zhang2014cross, o2014bridging}.
In order for a side-channel attack to be effective in practical scenarios for a security breach, it has to be executable without having physical access to the device being attacked~\cite{wakabayashi2017poster}. 
In the case of digital investigation, the investigator has the freedom to handle the device, and ideally, any investigative activity must not affect or change the digital information in the device~\cite{du2017processmodelsdfaas}.
Electromagnetic (EM) Side-channel Attacks is one approach that has shown promising results. It requires minimum physical manipulations to the device being inspected~\cite{hayashi2013efficient}. EM emissions of a device can be passively observed to infer both the internal operations being performed and the data being handled~\cite{sayakkara2018EMForensics}.
This condition is ideal for a digital investigator who attempts to ensure that the device does not go though any physical changes due to its investigation.
It is worth noting that hardware manufacturers are continuously trying to circumvent EM side-channel attack vulnerabilities through EM shielding and operation obfuscating enabled firmware.

This paper discusses the possibility for EM side-channel analysis as a potential case-advancing possibility for digital forensic analysis of IoT devices. A comprehensive analysis of the literature is provided identifying some promising avenues for research and their future potential. EM side-channel attacks for the recovery of cryptographic keys and other forms of important information are evaluated for potentially overcoming the encryption problem in digital forensics on IoT devices. Since the nature of EM emission phenomena is associated with the power consumption of computing devices~\cite{callan2015comparison}, the literature that focuses on power analysis attacks are also discussed where appropriate.


The contribution of this work can be summarised as follows:
\begin{itemize}

\item A comprehensive literature review and a comparative study of the research that has been carried out in EM side-channel analysis is provided and recent advances are summarised.

\item The scenarios where different EM side-channel attacks in the literature are relevant and applicable in digital forensic investigations are identified.

\item Light is shined on several new avenues of research that are possible to achieve in digital forensic investigations and cybersecurity through the adoption of EM side-channel analysis techniques.

\item The shortage of reliable tools and frameworks available to utilise EM side-channel analysis for digital forensic investigations on IoT devices is identified and the recommendations are made to overcome it. 

\end{itemize}

The rest of this paper is organised as follows. Section~\ref{side-channel-attacks} presents an overview of side-channel attacks. Sections~\ref{acquiring-em-emissions},~\ref{uniqueness-of-em-emissions}, and~\ref{information-leakage-from-em-emissions} explores approaches for acquisition, unique identification, and information leakage EM emissions relevant to digital forensics. In Section~\ref{standards-tools}, the advancements in wireless communication technologies and standardisation, and the legal background relevant to EM side-channels are discussed. Section~\ref{discussion} provides insights of possible future ethical directions of this technique. Finally, Section~\ref{conclusions} concludes the paper.

\section{Side-Channel Attacks}
\label{side-channel-attacks}

The topic of side-channel attacks spans a wide variety of techniques. Each side-channel attack on a computer system focuses on one specific unintentional leakage of information from either hardware or software~\cite{spreitzer2018systematic}. Some of such information leaking side-channels are listed below.

\begin{itemize}
\setlength\itemsep{-0.25em}
    \item The memory and cache spaces shared between different software.
    \item The amount of time a program takes to respond to different inputs.
    \item The sounds different components of computer hardware make.
    \item The amount of electricity a computer system draws.
    \item The EM radiation a computer hardware emits.
\end{itemize}

Computer programs contain conditional branches and loops in order to handle inputs and produce the intended output. Depending on the input values, the execution path of a program can differ, which may result in a different program execution time. It has been shown that the execution time of encryption algorithms can reveal information regarding the input values provided to it, which includes the encryption key~\cite{dhem1998practical}. For example, the square and multiplication segment in the Rivest–Shamir–Adleman (RSA) algorithm checks whether a key bit is 0 or 1 before moving into multiplication operations. Therefore, the observation of large number of execution times with the same key and different input data can lead to uncovering the key bits effectively~\cite{dhem1998practical, brumley2005remote, kocher1996timing}.

In environments where multiple virtual machines (VMs) run on the same hardware, such as cloud infrastructure, cache-based side-channel attacks are possible~\cite{zhang2014cross}. While each VM has its own virtual resources, many of them are mapped into shared physical resources including shared cache memories. It has been shown that an attacker running a VM on a virtualised environment can spy on a victim VM through the shared cache storage. This can lead to the extraction of sensitive information, including cryptographic keys~\cite{liu2015last}.

It has been shown that acoustic emanations from various components and peripherals of computer systems can be used to exfiltrate information~\cite{o2014bridging}. Genkin et al. showed that it is possible to distinguish between CPU operations by listening to acoustic emanations resulting in an attack on the cryptographic keys of the RSA algorithm~\cite{genkin2014rsa}.

Computer displays and their video cables have also been identified as an eavesdroppable EM source, which can leak the image being displayed on the display. Such leakages from CRT based displays have been known for several decades~\cite{van1985electromagnetic}. Video information provided to a computer display has synchronisation information to recognise between different lines of pixels and different frames, which are called horizontal and vertical synchronisations. By recognising this synchronisation information in the EM emissions, an attacker can reconstruct the images being displayed~\cite{hongxin2009recognition, elibol2012realistic}.

Kocher et al. were the first to introduce power consumption based side-channel attacks; \emph{simple power analysis} (SPA) and \emph{differential power analysis} (DPA)~\cite{kocher1999differential}.
SPA collects power consumption variation (in mA) over time with a high sample rate, such as twice the clock frequency of target cryptographic device.
The waveform of the power consumption, when plotted against time, contained patterns that corresponded to the instructions of the data encryption standard cryptographic algorithm (DES). If SPA can reveal the sequence of operations, it follows that this sequence depends on the data being handled by the algorithm (due to conditional branching). Designing code to minimise data dependent branching, which does not show characteristic power consumption patterns for specific operations, can prevent attackers from recognising what is executing on the device~\cite{zankl2018side}.

DPA is a technique that can be custom tailored for specific encryption algorithms. Kocher et al. used the DPA technique against DES~\cite{kocher1999differential}. The technique was able to guess the encryption key accurately, given sufficient cipher texts and power traces for those encryption operations. The authors claim that they have used DPA to reverse engineer various unknown algorithms and protocols on devices. The authors state that it may be possible to automate this reverse engineering process. Kocher et al. hints that these techniques (SPA, DPA) might be usable with EM emissions too in addition to power consumption.

While various side-channel attacks are possible on computer systems, it is possible to increase the advantages achievable by combining multiple side-channels that leak different kinds of information together~\cite{agrawal2003multi}. For example, power analysis and EM analysis can be performed together in order to reduce the errors and improve the accuracy of inferring the leaked information from a computer system.

\section{Unintentional Electromagnetic Emissions}
\label{acquiring-em-emissions}

EM radiation is the underlying technology for numerous of wireless communication. Meanwhile, it is a well documented fact that electronic devices generate EM radiation on unintended frequencies as a side effect of their internal operations~\cite{getz1996understanding}. Such unintended EM radiation are regulated by government agencies, such as \emph{Federal Communications Commission} (FCC) in the USA, due to the possible interference they can make on legitimate wireless communication and the potential health issues they can cause to the users of these devices. However, it is not possible to entirely avoid such emissions. Equipment manufacturers attempt to minimise it as much as possible~\cite{ott2011electromagnetic}.
This section discusses how EM signals are generated from different components of a computer system, what kind of information they may carry, and what types of methods and tools can be used to capture these signals.

\subsection{Hardware that Causes Electromagnetic Emissions}

As derived from Maxwell's equations, EM waves can be generated by electric currents varying over time. Characteristics of the EM waves being generated, such as frequency, amplitude, and phase, depends on the nature of the time varying electric current~\cite{maxwell1865dynamical}. Based on this principle, modern communication systems generate oscillating currents on antennas that generate EM waves that propagate over free space. They can be captured by another antenna with appropriate properties. Modern digital computer systems have a large number of components that depend on electric pulses or alternating currents for their operations. That leaves the space for EM waves to be generated at unexpected frequencies without the intention of the system manufacturer.

There are multiple computer components that operate in a coordinated, sequential fashion according to clock signals. Among them, both the CPU and RAM are of particular interest. The CPU performs a cycle of fetching instructions, decodes and executes them, while RAM maintains the data and corresponding instructions when a computing device is powered on. EM emission signals from these components contain a significant amount of side-channel information regarding the events related to software execution and data handling. On most IoT devices, the CPU and RAM are included in microcontroller (MCU) chips making it the most important EM source on-board.

\subsection{Sampling Electromagnetic Emissions}

The EM emission frequencies of a target device is unpredictable due to its dependability on various hardware characteristics. Therefore, it is difficult to have a universal purpose device that can be used to observe EM emissions from a target device and interpret side-channel information. It has been shown that small magnetic loop antennas can be used for the purpose of detecting EM emissions from computing devices~\cite{peeters2007power}.
When EM signals are captured by a loop antenna, it requires digital sampling before the data can be used for analysis. Theoretically, the sample rate of the equipment should be twice that of the maximum EM frequency required to be captured -- referred to as Nyquist frequency~\cite{smith1997scientist}. For this reason, EM signal sampling equipment must have a very high sample rate. The most commonly used equipment to capture EM signals are oscilloscopes and spectrum analysers with high sample rates. The digitised data these devices capture can be subsequently analysed in signal analysis software. However, access to such devices for information security professionals is not very common~\cite{wolfe2003setting}.

Software defined radios (SDRs) are getting increasingly popular among wireless hackers, hobbyists, and security enthusiasts who are interested in access to the radio frequency (RF) spectrum. An SDR consists of a minimal hardware component, which can be tuned to a range of RF frequencies and then digitise it with a fast analogue-to-digital converter (ADC). 
The processing of digitised RF data is handled entirely on software~\cite{tuttlebee2003software}. A wide variety of SDR hardware and software platforms are available~\cite{cass201340, ossmann2016software, ettus2015universal, blossom2004gnu}. Due to the enhanced flexibility provided by software, SDR platforms have become a perfect candidate for EM side-channel attack analysis research. A SDR can be used to scan through a wide range of frequencies to locate potential EM emissions from a computer system.

\begin{figure}[t!]
    \centering
    \includegraphics[width=0.55\textwidth]{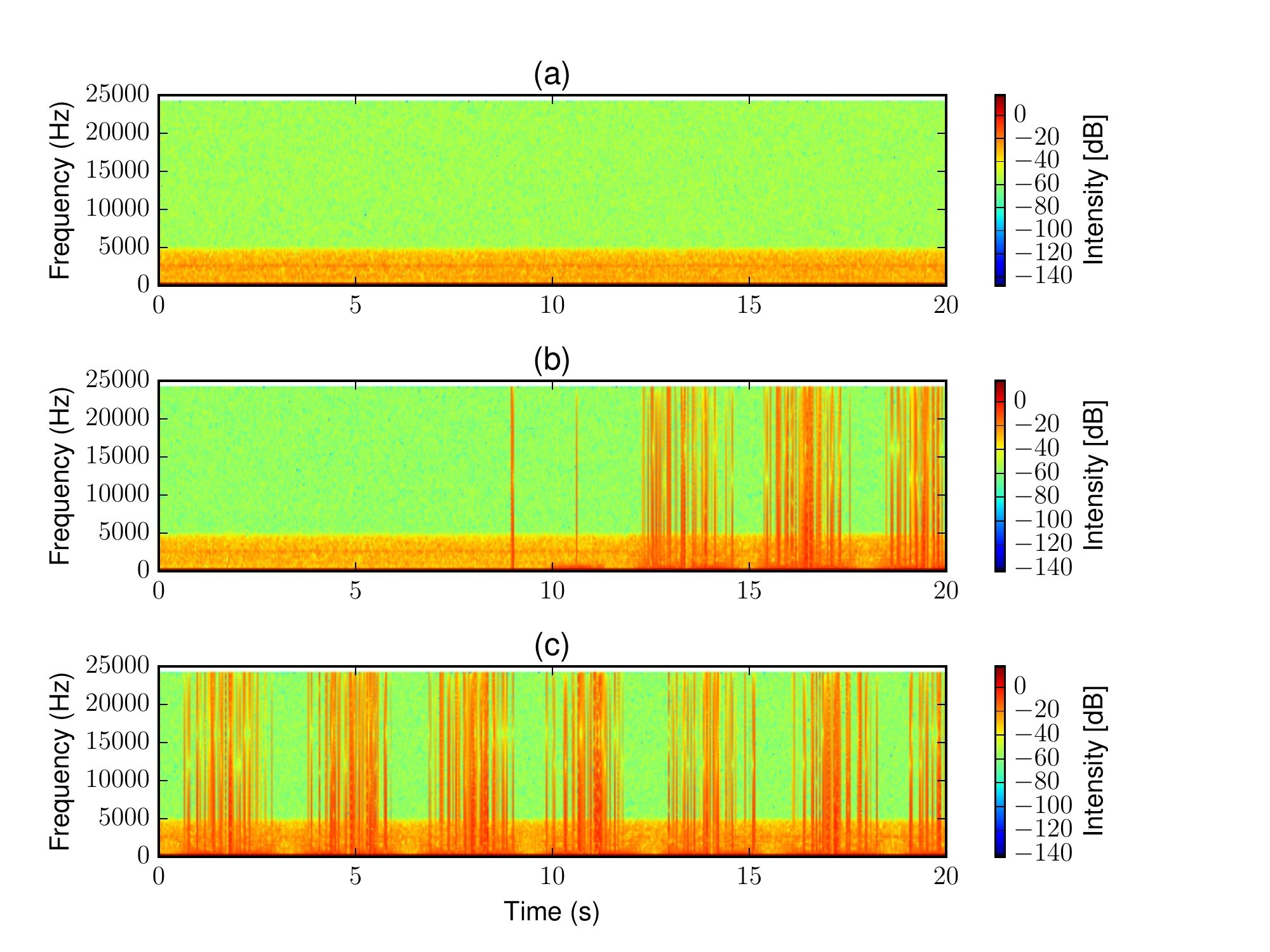}
    \caption{Spectrograms of AM demodulated EM emissions acquired from an Arduino device where (a) running a simple LED blink program, (b) reprogramming the device, and (c) running a complex LED blink program are depicted.
    }
    \label{fig:led-pwm-spectrogram}
\end{figure}

A simple set-up with an SDR platform can be used to demonstrate the unintentional EM signals.
An \emph{Arduino Leonardo} prototyping board is loaded with a simple program to blink an LED connected to it via the general purpose I/O pins. An antenna connected to an RTL-SDR dongle is placed close to the Arduino board in order to receive unintentional EM signals emitted from the board. The Arduino board consists of a microcontroller chip that operates at 16~MHz. However, the RTL-SDR dongle cannot be tuned to frequencies below 22~MHz. Therefore, a GNURadio script was programmed to tune the RTL-SDR dongle to the first harmonic of the Arduino clock, i.e., 32~MHz. Figure~\ref{fig:led-pwm-spectrogram} illustrates the spectrograms of three different EM signal samples gathered.

When two different LED blinking patterns are performed by two different programs separately, the Arduino emitted two completely different EM signal patterns. The spectrogram shown in Figure~\ref{fig:led-pwm-spectrogram}~(a) illustrates the EM signal observed from a simple LED blinking program, while the spectrogram in Figure~\ref{fig:led-pwm-spectrogram}~(c) illustrates the EM signal of a more complex LED blinking program. The spectrogram depicted in Figure~\ref{fig:led-pwm-spectrogram}~(b) shows the EM signal observed during the re-programming stage of the Arduino device from the first program to the second. 

\subsection{Connection Between CPU Instructions and Electromagnetic Emissions}

As a result of executing instructions in different combinations by the CPU, EM signal patterns are emitted at various frequencies and amplitudes.
Depending on the sequence of instructions, i.e., the exact program being executed, the output of EM noise from the CPU varies significantly. Due to this, systematically modelling and predicting possible EM signal characteristics of a computer processor is a difficult task. In order to identify unintentional EM emissions of a computer processor, the most practical method is scanning a large frequency spectrum for suspected EM signals and subsequently trying to interpret these identified signals for potential side-channel information. This arduous approach is a time consuming task that requires manual inspection by a human user. 

\begin{sloppypar}
In 2014, Callan et al. introduced a metric called SAVAT (Signal AVailability for an ATtacker) that measures the EM signal power emitted when a CPU is executing a specific pair of instructions (A and B). The authors show that different selections of A/B instruction pairs emit different SAVAT values, i.e., signal power~\cite{callan2014practical, zajic2014experimental}. An improvement to the SAVAT technique is a method called \emph{Finding Amplitude-modulated Side-channel Emanations} (FASE). The key idea behind the FASE technique relies on the phenomena that when a program activity is alternating at a frequency ($f_{alt}$) that affects any periodic EM signal originating from any source at a frequency $f_{c}$, it is possible to observe two side-band signals at $f_c - f_{alt}$ and $f_c + f_{alt}$ between the $f_c$ signal. Further improvements to SAVAT technique enabled the possibility of identifying both amplitude and frequency modulated EM emissions from CPUs~\cite{callan2015fase, prvulovic2017method, yilmaz2018capacity}. While it is evident from existing studies that EM side-channel leakage is available across various type of CPUs, further studies are necessary to identify the effect of different CPU architectures to the produced EM emissions.
\end{sloppypar}

\section{Electromagnetic Emissions as a Signature}
\label{uniqueness-of-em-emissions}

When a computing device running a program generates EM emissions, the patterns observable depend on the precise settings of the device.
In the EM emission spectrum of the Arduino device in Figure~\ref{fig:led-pwm-spectrogram}, it is clearly evident that both the hardware and software settings have influenced the EM emission patterns.
The signal captured at the first harmonic frequency of the Arduino device's system clock, i.e., 32~MHz, is showing a varying patterns in the spectrogram view according to the changes made to the device. Therefore, it is clear that the target device's system clock is the main source of EM radiation. 
The design of the printed circuit board (PCB), and characteristics of the electronic components provide variations to this strong signal. Meanwhile, subfigures (a), (b), and (c) of Figure~\ref{fig:led-pwm-spectrogram} clearly show that the instruction sequence, i.e., the program being executed on the CPU, has a significant influence to the EM emission pattern.

\subsection{Electromagnetic Emissions as a Hardware Signature}

Despite the software components available on a computing device, it is important to investigate whether the hardware alone can provide a recognisable EM emission pattern.
Such a capability can lead to profiling of hardware devices and components uniquely.
It has been shown that a simple EM signal acquisition device called RTL-SDR, which demonstrated the capability to capture EM emissions from an Arduino device, can be used to profile computing devices uniquely. Laput et al. used a similar device to acquire EM signals that were successfully applied to a support vector machine (SVM) classifier to uniquely distinguish the EM source device~\cite{laput2015sense}. This possibility has led to the idea that EM emissions from an electronic device owned by a person can be used as an authentication token of the person instead of relying on conventional methods, such as Radio Frequency Identification (RFID) tags~\cite{bianchi2016wearable, yang2016id}.

This uniquely distinguishable EM emission patterns of a known electronic device can help to identify any potential alteration that may have applied to it. For example, a known electronic device can be altered at the hardware fabrication level for malicious purposes, such as accessing stored data or eavesdropping on users' activities. Such hardware modifications result in a changed EM emission pattern that can be used to identify it~\cite{yang2017exploiting}. Similarly, a genuine electronic device can be replaced by a counterfeit electronic device for a malicious objective. It has been shown that even when counterfeit hardware attempts to follow the design of the genuine device, it still creates distinguishably different EM emission patterns compared to their original product~\cite{ahmed2017radiated}.

\subsection{Electromagnetic Emissions as a Software Signature}

When software runs on different computing devices, it is clear that the hardware EM emissions are influenced by the software instructions being executed. It is important to consider this influence from two different aspects. The first aspect is how uniquely the EM emissions of different software running on the same hardware platform can be recognised. This can be used to pin point to the exact software running on a device. The second aspect is how unique the same software program is when it is running across various hardware platforms. It facilitates the unique detection of a specific piece of software.

When software systems are being developed, requirements arise to debug their behaviour or find ways to increase their performance. Instrumenting software by applying logging events and break points are the most common ways to identify where complex software is not performing as expected. These developmental options affect the performance of the software being inspected in addition to the overhead of their placement each time a copy of the software needs to be inspected. It has been shown that unintended EM emissions of the CPU can be used to inspect software execution sequences without having to instrument the software~\cite{callan2016zero,callan2016analyzing, han2017watch, espitau2017side}. Even when the same program is running on different devices, the ability to identify the instruction execution sequence can help to uniquely identify the software itself.

The capability to detect software code execution sequence has opened up the opportunity to identify when a computing device is running software code not intended by the manufacturer or the owner. One possible scenario can be software bugs or hardware faults that cause an IoT device to execute unexpected instruction sequences. Another possible scenario can occur when an IoT device is under an attack causing it to run malware or an unintended part of the device's genuine software. Stone et al.~\cite{stone2015radio, stone2015detecting} and Nazari et al.~\cite{nazari2017eddie} showed that such abnormal deviations of software code executions on computing devices can be detected using the corresponding EM emission patterns from its execution.

Even though it has been identified that EM emission patterns are associated with both the hardware and software characteristics of the source device, the format of a captured EM signal used for this identification can vary. Instead of directly using time-domain EM signal traces, one such alternative format is RF-DNA fingerprinting. This is a technique to fingerprint the physical layer of RF transmitting devices, which includes WiFi, Bluetooth, Zigbee, GSM devices, and even RADAR antennas. This technique  has been used to identify rogue devices in a deployment using their RF signals without physically inspecting them~\cite{reising2012exploitation, dubendorfer2013using, danev2012physical, lukacs2015rf}. Deppensmith et al. showed that RF-DNA technique can be applied to unintentional EM emission fingerprinting on computing devices reliably~\cite{deppensmith2014optimized}. However, the evaluations performed by Stone et al. on microcontroller based IoT devices indicates that further studies are necessary to conclude the most reliable format to represent unintentional EM signals~\cite{stone2016comparison}.

\begin{figure}[t!]
    \centering
    \includegraphics[width=0.5\textwidth]{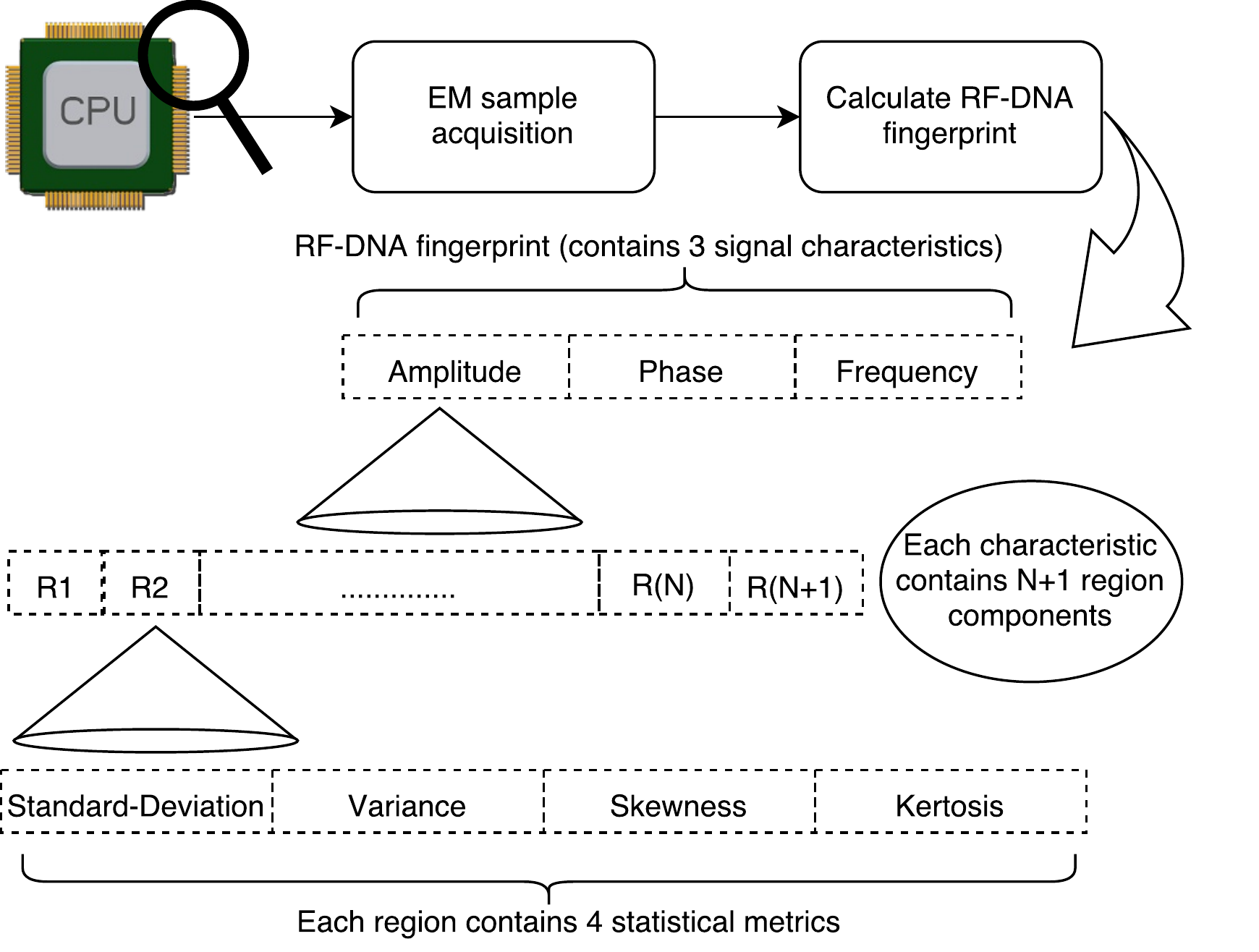}
    \caption{The RF-DNA fingerprinting process.
    }
    \label{fig:rf-dna-structure}
\end{figure}

Figure~\ref{fig:rf-dna-structure} illustrates the structure of an RF-DNA fingerprint. When calculating it, the EM signals emitted from a device on a selected frequency is captured, filtered, and amplified appropriately to achieve a clean trace. From this acquired time-domain EM trace, the three signal characteristics; \emph{Amplitude}, \emph{Phase} and \emph{Frequency}, are separately considered for further processing. Each signal characteristic is broken into $N$ equally sized regions and then for each region, four statistical metrics; \emph{standard deviation}, \emph{variance}, \emph{skewness}, and \emph{kurtosis} is calculated.
The standard deviation and variance metrics measure the spread of data samples, while the skewness and kurtosis metrics measure the symmetry and the sharpness of a data sample.
Furthermore, the signal itself is again considered as a one entire region, i.e., the $(N+1)^{th}$ region, to calculate the same statistical metrics. As Figure~\ref{fig:rf-dna-structure} illustrates, each of these calculated statistical metrics are arranged in a single vector, which becomes the RF-DNA fingerprint of the originally acquired EM trace from the device.

\section{Information Leaking Electromagnetic Emissions}
\label{information-leakage-from-em-emissions}

This section dives into the question of what information is contained in an EM emission trace of a particular computing system. From a digital forensic perspective, both the kind of software running on IoT devices and the data being handled by each software application are potentially of significant interest. Even if EM side-channel analysis cannot reveal all of data being handled by an IoT device platform, extracting critical information, e.g., cryptographic keys, can help progress forensic analysis.

\subsection{Observable Electromagnetic Spectrum Patterns}

While there exists a wide variety of microcontroller chips used on IoT devices, Sohaib et al. has shown that it is still viable to perform EM side-channel attacks on them~\cite{sohaib2016side}. When considering information leakage from an EM emission trace,
visually inspecting the time-domain signal is the first observational technique. This approach is called simple electromagnetic analysis (SEMA), which evolved from the simple power analysis (SPA) introduced by Kocher et al.~\cite{kocher1999differential}. Another way of performing visual observations is by transforming the EM trace into the frequency domain and plotting it as a spectrogram. This enables observation of different signal patterns distributed over multiple frequencies.

Multiple published works have demonstrated the effectiveness of the SEMA approach in extracting critical data from computers, including cryptographic keys. The El Gamal and RSA algorithms implemented using GnuPG library were attacked by observing critical CPU operations~\cite{genkin2015stealing}. Furthermore, elliptic curve based cryptographic algorithms (ECC), such as Elliptic Curve based Diffie Hellman (ECDH) and Elliptic Curve based Digital Signature Algorithm (ECDSA), are identified to be vulnerable to EM side-channel attacks with the SEMA approach~\cite{goubin2003refined, genkin2016ecdh, genkin2016ecdsa}. Due to the low computational overhead in ECC algorithms, many mobile devices and IoT platforms tend to employ ECC algorithms to secure data. This indicates that such devices can be inspected through EM side-channels to access cryptographically protected data.

\subsection{Differential Electromagnetic Analysis (DEMA)}

If a large number of EM traces from a computing system can be observed executing specific software, it is possible to identify the data bits involved in the operations that appear across the large number of EM traces. While being impractical to perform under real-world attack scenarios due to the challenge of collecting such a large number of EM traces from a target computer, this may be the only resort for EM side-channel analysis when it is not possible to extract information from visual observation of a single EM trace. 

DEMA, a variant of Differential Power Analysis (DPA), uses the variation of EM emissions of a CPU to discover variables used in an executing program, such as encryption algorithms~\cite{kocher1999differential, kocher2011introduction}. When a bit in a CPU register is flipped from 0 to 1 or vice versa, it consumes an amount of energy, which is reflected in the corresponding EM emission. Some CPUs may emit a higher EM signal when switching a register bit from 0 to 1 than vice versa since that operation can lead to a higher energy consumption~\cite{peeters2007power}. Due to this, when the contents of a complete CPU register are modified, it is possible to identify the \emph{hamming distance} between the previous and new state using the resulting EM emission. Since every instruction running on a CPU affects the values on different registers, this means that attackers can identify instructions being executed and intermediate variables used based on the EM observation.

\begin{figure}[t!] 
    \centering
    \includegraphics[width=0.48\textwidth]{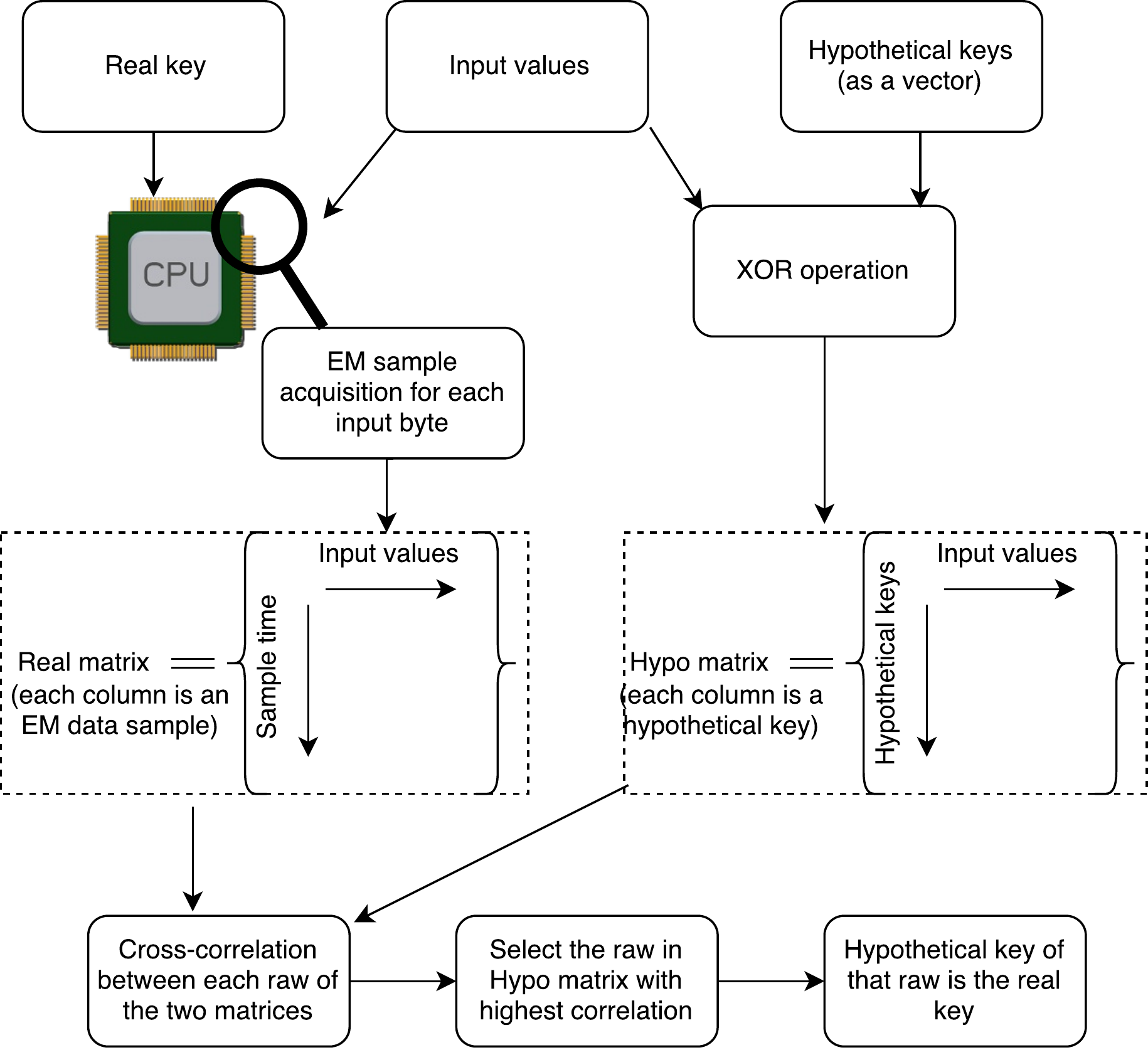}
    \caption{EM analysis of XOR-Cipher algorithm to extract the encryption key.
    }
    \label{fig:em-analysis}
\end{figure}

Figure~\ref{fig:em-analysis} illustrates how differential EM analysis is used to identify the key of a simple XOR-cipher. Initially, a large set of input data bytes and all possible key bytes are used to perform XOR-operations and the hamming distances of the resulting values are stored in a matrix called \textit{Hypothetical Matrix}, as shown in Figure~\ref{fig:em-analysis}. The objective is to find the hypothetical key of the matrix that generates matching hamming distances for the same input values. For this, the input values are fed to software running on the CPU, where the XOR operation is performed with the unknown key. EM emissions are sampled for each input value to generate the \textit{Real Matrix}, where each column contains an EM signal sample for its corresponding input. Calculating cross-correlations between rows of the two matrices can find the corresponding row in the \textit{Hypothetical Matrix} that provides the best correlation of hamming distances. The hypothetical key corresponding to this row is likely the encryption key used for the XOR operation inside the CPU.

Standard encryption algorithms, such as the Data Encryption Standard (DES) and the Advanced Encryption Standard (AES), employ XOR operations at various stages in their functionality using chunks of the encryption key and input data. Therefore, DEMA attacks are possible by attacking each chunk of the key being used with the XOR operations. Such an attack reveals parts of the encryption key, which have to be combined at the end. However, in a real-world setting, the attacker may not have enough EM emission samples of the encryption operations to calculate the correct part of the key used. This results in lists of possible key chunks for each segment of the encryption key used in the algorithm. The problem of identifying the correct parts of the key to build the complete encryption key is called the \emph{Key Enumeration Problem}, which can be solved within a reasonable computational overhead~\cite{bogdanov2015fast}.

Quisquater et al. practically demonstrated that EM analysis is a viable option to the aforementioned power analysis attack on computer CPUs~\cite{quisquater2001electromagnetic}. By precisely moving the EM probe over a microcontroller, the authors were able to build an accurate 3-dimensional EM signature of the chip running an idle loop. It was shown that the radiation spectrum of each processor was sufficiently unique to use as a distinguishable feature for processor identification. These experiments were performed in a Faraday cage to minimise the external noise effects and the EM emissions were captured using a small magnetic loop antenna (diameter $\approx$ 3mm). An oscilloscope digitised the signal for analysis. Gandolfi et al.~\cite{gandolfi2001electromagnetic} applied DEMA to extract encryption keys from three different chips used on smartcards namely; COMP128, DES, and RSA. According to this study, the SNR of EM emissions from these chips is higher than the SNR of power consumption analysis. This results in the extraction of more information from the DEMA technique as compared to simple power analysis attacks~\cite{messerges1999investigations}.
Asymmetric key encryption, such as RSA, can also be attacked by identifying the individual modular exponentiation operations performed within the algorithm through EM emissions~\cite{witteman2011defeating}.

Due to the modern electronic multimedia distribution model, e.g., music, movies, and ebooks, end-users can often become the attackers. These users might have the malicious intention of breaking encryption or sharing Digital Rights Management (DRM) protected data. Since the victim device is owned by the attacker, unlimited physical access to the hardware and software becomes available for the attacker through various side-channel attacks. White Box Cryptography (WBC) was introduced as a solution to this; whereby cryptographic algorithms and keys are combined with random codes and random data to create an obfuscation that makes side-channel attacks more difficult. However, it has been shown that EM side-channel attacks, such as DEMA, are still capable of extracting encryption keys despite of the application of WBC techniques~\cite{sanfelix2015unboxing}.

Recently, Camurati et al. made an important discovery that extended the previously known capabilities of EM side-channel analysis of cryptographic operations on IoT devices~\cite{camurati2018screaming}. It was shown that mixed-signal processors, such as system-on-chips (SoCs), that contains a radio transceiver and a CPU on the same silicon die, can cause long distance EM leakages. This occurs when the CPU noise gets modulated into the radio transceiver's emission -- extending the range of the CPU EM side-channel. As the usage of SoCs is getting increasingly popular on IoT devices, this latest type of EM side-channel leakage, called \emph{screaming channels}, has significantly increased the potential attack surface.

\subsection{Analysis on Wireless-Powered Devices}

Unlike traditional computing devices that have their own power source to run CPU operations, wireless-powered devices, e.g., passive RFIDs, depend on an external RF field provided by the device's reader for power~\cite{calari1997rfid}. IoT devices are ideal candidates to be powered by wireless means. EM side-channel analysis on such devices is challenging due to the presence of a strong RF field from the reader, which obfuscates the weak EM emissions of the devices themselves. However, RFID based devices are being used in critical systems, such as secure access control to buildings and electronic payments, where cryptographic operations are performed on-board. Therefore, investigating the EM side-channel capability on such devices is important from both security and forensic standpoints.

Hutter et al. demonstrated the capability to perform EM side-channel analysis on RFID based devices using a custom-made RFID tag as a proof of concept~\cite{hutter2007power}. Using this custom set up, the authors were able to recover the AES key used in a challenge response protocol between the RFID tag and the reader. In order to avoid the disturbance from the RF field of the reader device, the RFID circuitry was placed outside the reader's RF field, while the power harvesting antenna is kept inside the reader's RF field. The two components were connected through a sufficiently long wire. This enabled the measurement of the EM emissions from the RFID circuitry without interference. However, it is not possible to follow a similar approach in a regular RFID tag as the antenna and RFID circuitry are inseparable by any reasonable means.

Kasper et al. performed EM side-channel attacks on RFID based smart-cards in a more realistic setting. This research employed commercially available smart-cards and performed the attacks within the RF field of the RFID reader~\cite{kasper2009side}. When the RFID smart-card is consuming more energy, the amplitude of the RFID reader’s RF field becomes lower. Similarly, when the RFID smart-card is consuming less energy, the amplitude of the RFID reader’s field is higher. This means, the power consumption of the RFID tag is reflected in the amplitude of the RFID readers carrier frequency. 
Kasper et al. used this signal as the EM side-channel to attack internal operations of the RFID reader. A computer-controlled USB-oscilloscope and a computer-connected custom-made RFID reader was used to attack the RFID tag while capturing EM fluctuations using a small RF loop probe. This set up was used to perform a \emph{correlation power analysis} (CPA) attack to extract the symmetric keys used in DES and 3-DES implementations on the smart-card successfully.

Recent research by Xu et al. demonstrated that RFID based smart-cards that employ side-channel attack mitigation techniques, such as \emph{head and tail protection}, are not effective enough against EM side-channel analysis attacks~\cite{xu2018side}. In their work, encryption keys used for the 3DES algorithm were demonstrated to be recoverable. In light of this attack vector, it is important to note that wireless-powered IoT devices are also susceptible to threats from EM analysis based attacks.

Many smart-card-based fraud, such as stealing credit card details, involve malicious devices, such as \emph{card skimmers}, that can read and store data from the cards. Investigators face the challenge of identifying victims of such skimming devices due to the fact that card details are encrypted when stored on these devices. Souvignet and Frinken demonstrated that correlation power analysis, which is a variant of DPA attack, can be used to extract the details of victim smart-cards from such skimmer devices by physically tapping into a seized device~\cite{souvignet2013differential}. The success of their work indicates that EM side-channel analysis can be even more promising in extracting such evidence without requiring any physical alterations to the device itself.

\subsection{Countermeasures to Electromagnetic Side-Channels}

As EM side-channel analysis has been shown to be successful on recovering data from computing devices, various countermeasures have been explored to counteract it on both software and hardware levels~\cite{zankl2018side}. Masking variables by using random values alongside the operations is a basic software-based countermeasure that has been proven to not be effective enough against EM side-channel attacks~\cite{kim2008differential, chari2002template}. Various other approaches including the randomisation of the operation sequences or lookup tables of algorithms~\cite{saputra2003masking, kim2016protecting}, avoiding instructions pairs executing adjacently that are known to emit distinguishable EM patterns~\cite{callan2014practical, zajic2014experimental}, and accessing critical data using pointers instead of values~\cite{witteman2008secure} require further studies to see how effective they are against EM side-channel attacks.

Quisquater et al. suggests several hardware design countermeasures to these attacks~\cite{quisquater2001electromagnetic}. Actions that can be taken by hardware designers includes minimising metal parts in a chip to reduce EM emissions, the use of Faraday cage like packaging, making the chip less power consuming (which leads to less unintentional emissions), asynchronism (i.e., design the chip not to use a central system clock and instead operate asynchronously), and the use of dual line logic (i.e., using two lines that in combination of two bits represents a state instead of a single line that simply represent 0 or 1 states). Furthermore, it has been shown that it is possible to mathematically model an electronic chip during the design phase to identify and avoid potential information leakages through EM side-channels~\cite{ishai2003private, standaert2009unified}.

\section{Standards and Tools}
\label{standards-tools}


\begin{figure*}
    \centering
    \includegraphics[width=0.85\textwidth,trim={0 0 0 2.5cm},clip]{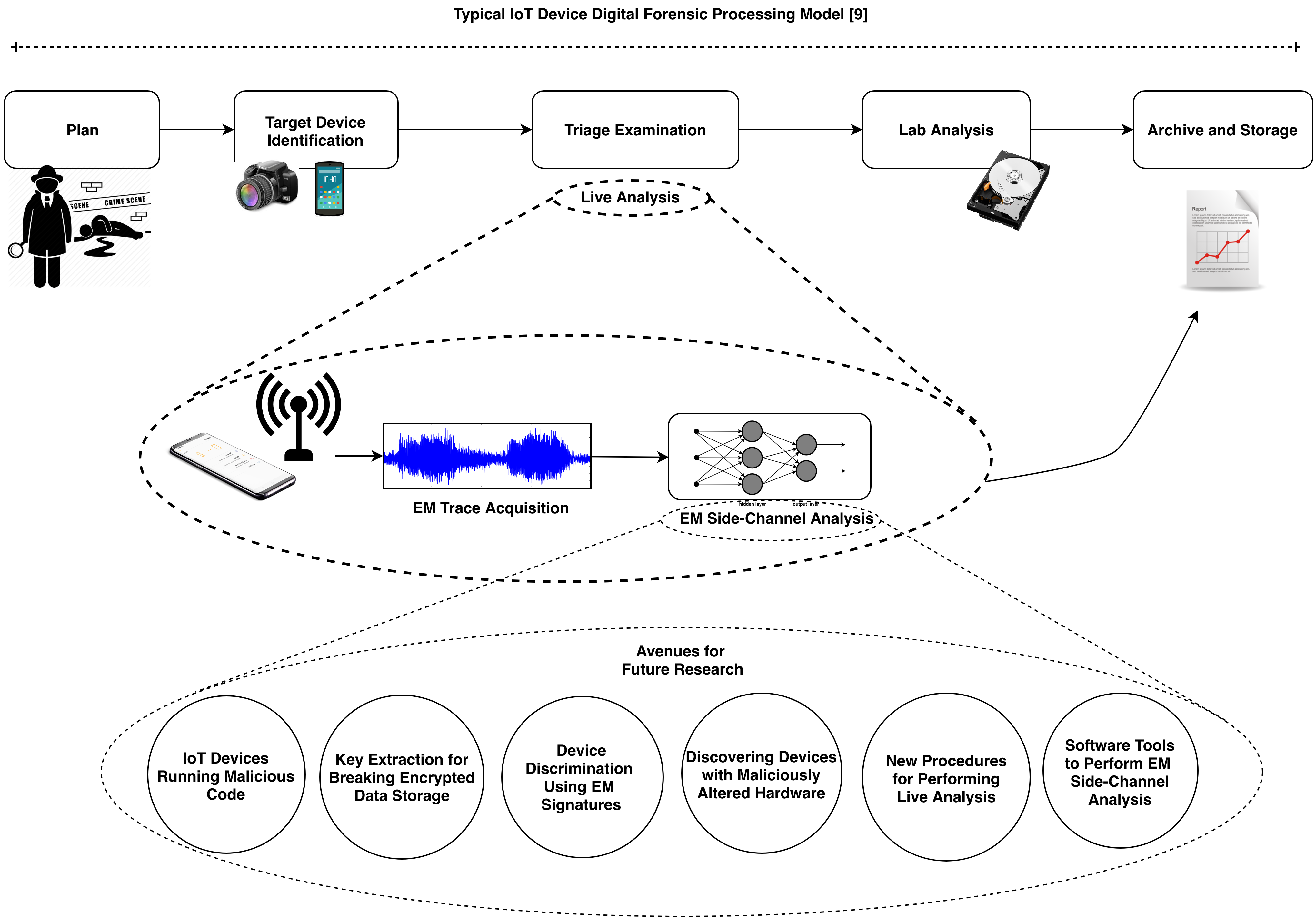}
    \caption{Altering the traditional IoT digital evidence acquisition process~\cite{du2017processmodelsdfaas}, EM side-channel analysis can help in various ways for live analysis of computing devices.}
    \label{fig:em-forensic-process}
\end{figure*}

EM side-channel attacks are not currently commonly being used for digital forensics purposes. Therefore, it can be too early to find any existing standards or tools on EM side-channel analysis for digital forensics. However, in order for future establishment of standards and tools, it is important to review the relevant standards and tools in both hardware and software security domains.  

The concerns of electromagnetic wave emissions from IoT devices from the software perspective are mostly concentrated towards the wireless communication technologies, such as WiFi, Bluetooth, and proprietary IoT protocols, e.g., Zigbee~\cite{golmie2003bluetooth, iyer2015detecting}. Meanwhile, unintentional EM emission minimisation is generally left to those involved in the hardware design and manufacturing process. The term \emph{electromagnetic compatibility} (EMC) refers to a device's unintentional EM emissions that can affect the functionality of other devices and the health of humans who are exposed to it~\cite{getz1996understanding}. The Federal Communications Commission (FCC), the Food and Drug Administration (FDA), the International Electrotechnical Commission (IEC), and the European Union (EU) are examples of authorities concerned with EMC~\cite{barron2007creating, hayashi2016state}. However, regarding the question of EM side-channel information leakage from general purpose electronic devices, there are no such rules to govern the manufacturers. Instead, only guidelines exist, which may or may not be followed~\cite{ISOIEC17825,ISOIECTS30104}.

Once a hardware device's design is completed and manufacturing commences, it is a challenging task to apply mitigation steps should the EMC tests reveal that it does not meet requirements. In the worst case scenario, the minimisation of EM emissions may require a complete reworking of the PCB used in the device or a replacement of a critical electronic component. Due to the potential for costly manufacturing disruption, the minimisation of EM emissions is necessary from the initial hardware design phase. Due to the fact that EM side-channel information leakage is not a problem limited to hardware manufacturers, a joint effort by both hardware and software developers to establish standards is necessary.

In order to facilitate the assessment on EM side-channel security threats, standardised tools and frameworks are highly necessary. \emph{Test vector leakage assessment} (TVLA) is a technique that can be used to assess the resistance of cryptographic implementations against hardware side-channel attacks~\cite{becker2013test}. TempestSDR is a software tool that can be used with a large variety of hardware platforms, e.g., the universal software radio peripheral\footnote{\url{https://www.ettus.com/}} (USRP) or HackRF \footnote{\url{https://greatscottgadgets.com/hackrf/}}, to eavesdrop on computer monitors by capturing the EM signals emitted by the video cables~\cite{tempestsdr, Sayakkara:2018:AEE:3230833.3234690}. Multiple commercial and open source products exist that can be used to break encryption on microcontroller based IoT devices, such as ChipWhisperer~\cite{chipwhisperer,o2014chipwhisperer} and Riscure Inspector~\cite{riscure, RamsayTempest}. Blanco et al. presented a side-channel trace acquisition framework called SCAP, which is targeted at general purpose computing devices (including mobile devices). While the framework does not currently perform side-channel attacks, the objective is to provide a platform to build future analysis tools~\cite{blanco2017framework}.
Such tools enable IoT system developers to test the robustness of their hardware against physical side-channel attacks and identify information leakage.

\section{Discussion}
\label{discussion}

Having discussed the scientific literature related to EM side-channel analysis attacks, it is important to identify the future impact it may cause in the domain of digital forensics on IoT devices. This section highlights some of the potential ways this impact may occur in the future under different themes.
Figure~\ref{fig:em-forensic-process} illustrates the avenues for future research in this direction.
Many of these future potentials are already starting to be realised and others are ambitious predictions that can prove significantly beneficial to digital forensics.

Recent versions of mobile operating systems, e.g., Android and iOS, secure their internal storage using encryption. Critical information necessary for digital forensic investigation can be inaccessible due to being stored in an encrypted form~\cite{casey2008impact, zdziarski2008iphone, hoog2011android}. This includes encrypted emails, encrypted instant messenger applications, encrypted files, and encrypted storage partitions.
While the increasing application of cryptographic protection on computing devices poses a challenge to traditional digital forensics, it can open up new opportunities to EM side-channel attacks~\cite{lillis2016challenges}. EM side-channel attacks require a large number of traces acquired from a victim device while the device is performing cryptographic operations using a single key. The most common encryption operations occurring on older systems are the secure socket layer (SSL) based web traffic~\cite{fahl2013rethinking}. With encrypted data storage becoming commonplace, EM side-channel attacks can potentially be performed by observing cryptographic operations during live data forensic analysis procedures~\cite{hay2009live}.

Instead of using a single side-channel attack in isolation, combinations of multiple side-channel attacks directed towards a single computer system can prove more fruitful. It has been proven that power and EM side-channel analysis can be combined to achieve better results~\cite{agrawal2003multi}. There can be some operations of the CPU that are more clearly reflected in the device's power consumption than in the EM emission and vice versa. Similarly, malware running on a victim computer can aid an EM side-channel attacker to extract additional information (over the EM side-channel alone) by intentionally modulating data into the EM emission of the CPU or the monitor~\cite{cheddad2010digital, yang2017comm, Sayakkara:2018:AEE:3230833.3234690}.

The unintentional EM emissions from computing devices can cause interference to other radio signals in the vicinity. This phenomena is evident in laptop computers, which have been shown to modulate signals from commercial AM radio stations~\cite{macbookeminterference}. IoT devices already use this interference phenomena to communicate purposefully with other devices by modulating the ambient RF signals. This is called \emph{backscatter} communication technology~\cite{liu2017backscatter}. There are various carrier wave sources that have been tested in the literature including TV transmission stations and WiFi access points~\cite{liu2013ambient, kellogg2014wi, bharadia2015backfi, zhang2016hitchhike}. The potential of using this backscatter phenomena to eavesdrop on internal CPU operations of IoT devices by listening to ambient RF sources warrants further exploration.

Devices deployed in wired networks, such as routers and switches, are known EM noise sources. It has been already showed that MAC addresses in Ethernet frames can be extracted by performing SEMA analysis on the EM emissions from wired routers~\cite{schulz2016trust}. When it is required to perform an investigation on a live wired network, it is necessary to be connected in order to inspect packets~\cite{corey2002network}. In situations like this, the EM emissions of routers and switches might be able to provide an approximate picture of the workload and traffic on the network~\cite{goudos2008emi}.

Recent advances that have been made in the area of artificial intelligence (AI) (incorporating machine learning (ML) and deep learning (DL)) have demonstrated promising applications to many other domains across computer science. Various tasks where human intuition was required to perform decision making are now being replaced with ML/DL powered algorithms. Software libraries and frameworks are becoming increasingly available in order to assist the building of applications that have intelligent capabilities. Examples include the automated detection of malicious programs~\cite{kolter2004learning}, image manipulation ~\cite{saboia2011eye}, and anomaly detection in network traces~\cite{mukkamala2003identifying}.

EM side-channel analysis techniques 
that previously required human intervention can be automated through the development of AI algorithms. Recent work by Wang et al.~\cite{wang2018deep} applied deep learning algorithms including multi-layer perceptron (MLP) and long short-term memory (LSTM) to detect anomalies in the code of simple IoT devices, e.g., Arduino and Raspberry Pi, through the power consumption side-channel. Therefore, it is potentially possible to extract better information from EM traces than the current manual observations are capable of achieving. Several examples that were discussed in previous sections already leverages AI techniques to recognise EM trace patterns, which strongly hints the future role that can be played by AI algorithms in EM side-channel analysis for digital forensics~\cite{laput2015sense,lerman2011side,callan2016zero,callan2016analyzing,nazari2017eddie,stone2016comparison}.

\section{Conclusions}
\label{conclusions}

Traditionally, digital forensics focuses on analysing traces left behind by suspects on digital devices by inspecting file storage, log files, network traces, etc. Live data forensics can also be performed on systems that require more sophisticated investigative techniques and skills. As computing systems transform from less privacy and security concerned platforms into hardened platforms that are designed with security in mind from their inception, the typical work conducted by digital forensic investigators must change accordingly. Cryptographically protected storage systems is one of the largest challenges hindering efficient digital forensic analysis. EM side-channel analysis has been demonstrated as a potential door-opener for cryptographically protected data storage and communications from a security perspective, which can be built upon and adopted for digital forensic purposes.

This paper comprehensively analysed the literature on EM side-channel attacks with the goal of applying the technique to assist digital forensic investigations on IoT devices. While various mitigation techniques have been suggested and applied to counter against EM side-channel attacks, existing literature demonstrates that such attempts have not been successful in reducing the prevalence of this attack vector.
EM side-channel analysis is still in its infancy for digital forensic applications, which demands court-admissible, forensically-sound processing when used not only for obtaining security keys but also for the detection of unintentional data leakage.
However, this technique has significant potential to have a substantial impact on the field and enable the progression of otherwise stalled investigative cases involving both IoT devices, and encrypted computing devices in general.

\begin{table*}[t]
\centering
\caption{Categorisation of the Literature on EM Side-Channels.}
\label{tab:summary-table}
\renewcommand{\arraystretch}{1.6}
\begin{tabular}{|l|l|l|}
\hline
\rowcolor[HTML]{EFEFEF} 
\multicolumn{1}{|c|}{\cellcolor[HTML]{EFEFEF}\textbf{Objective}} & \multicolumn{1}{c|}{\cellcolor[HTML]{EFEFEF}\textbf{Technique}} & \textbf{References}                                                                                                                                                                                                                                                                   \\ \hline
                                                                 & SAVAT                                                           & \cite{callan2014practical, zajic2014experimental, prvulovic2017method}                                                                                                                                                                                                              \\ \cline{2-3} 
                                                                 & FASE                                                            & \cite{callan2015fase, prvulovic2017method}                                                                                                                                                                                                                                          \\ \cline{2-3} 
                                                                 & SDR Signal Acquisition                                          & \cite{tuttlebee2003software, cass201340, ossmann2016software, ettus2015universal, blossom2004gnu}                                                                                                                                                                                   \\ \cline{2-3} 
\multirow{-4}{*}{Acquisition of EM Emissions}                    & General                                                         & \cite{getz1996understanding, ott2011electromagnetic, jabbar1991radio, peeters2007power, sohaib2016side}                                                                                                                                                                             \\ \hline
                                                                 & Time/Frequency/Hilbert Transform Domains                    & 

\begin{tabular}{@{}l@{}}\cite{yang2016id, stagner2013detecting, ahmed2017radiated, stone2015radio, stone2015detecting, callan2016zero, callan2016analyzing} \\ \cite{nazari2017eddie, clark2013current} \end{tabular}
                                               
                                                                 \\ \cline{2-3} 
                                                                 & RF-DNA                                                          & \cite{laput2015sense, reising2012exploitation, dubendorfer2013using, lukacs2015rf, deppensmith2014optimized}                                                                                                                                                                        \\ \cline{2-3} 
\multirow{-3}{*}{EM as a Signature}                              & General                                                         & \cite{bianchi2016wearable, yang2017exploiting, stone2016comparison} 

\\ \hline
                                                                 & SEMA / Spectrum Observation                                     & 

\begin{tabular}{@{}l@{}}\cite{hayashi2013efficient,van1985electromagnetic, elibol2012realistic, kocher1999differential, agrawal2003multi, genkin2015stealing, genkin2015get, genkin2016ecdh} \\ \cite{genkin2016ecdsa, belgarric2016side} \end{tabular}  
                                                                 
                                                                 \\ \cline{2-3} 
\multirow{-2}{*}{Cryptographic Key Extraction}                   & DEMA                                                            & 


\begin{tabular}{@{}l@{}}\cite{kocher1999differential, peeters2007power, kocher2011introduction, bogdanov2015fast, quisquater2001electromagnetic, gandolfi2001electromagnetic, messerges1999investigations, witteman2011defeating} \\ \cite{sanfelix2015unboxing, hutter2007power, kasper2009side, chari2002template,camurati2018screaming} \end{tabular}

\\ \hline
                                                                 & Hardware Countermeasures                                        & \cite{ fritzke2012obfuscating, quisquater2001electromagnetic, ishai2003private, standaert2009unified}                                                                                                                                                                                                        \\ \cline{2-3} 
\multirow{-2}{*}{Countermeasures for EM Side-Channels}           & Software Countermeasures                                        & \cite{callan2014practical, zajic2014experimental, kim2008differential, chari2002template, saputra2003masking, kim2016protecting, witteman2008secure, standaert2009unified}                                                                                                          \\ \hline
Tools and Frameworks                                            & Various                                          & 

\begin{tabular}{@{}l@{}}\cite{cass201340, ossmann2016software, ettus2015universal, blossom2004gnu, tempestsdr, chipwhisperer,o2014chipwhisperer, riscure} \\ \cite{RamsayTempest, blanco2017framework} \end{tabular}

\\ \hline
\end{tabular}
\end{table*}


\section*{APPENDIX}

A categorisation of the literature of EM side-channel attacks and related areas are listed in Table~\ref{tab:summary-table}.

\section*{References}

\bibliography{mybibfile}

\end{document}